\documentclass[]{emulateapj}
\shorttitle{Planetary System Around HD\,155358}
\shortauthors{Cochran et al.}
\begin{document}
\slugcomment{To Appear in The Astrophysical Journal, 1 September 2007 issue}

\title{A Planetary System Around HD\,155358: \\
    The Lowest Metallicity Planet Host Star
\footnote{Based on observations obtained with the Hobby-Eberly Telescope,
which is a joint project of The University of Texas at Austin, the
Pennsylvania State University, Stanford University,
Ludwig-Maximilians-Universit\"{a}t M\"{u}nchen, and
Georg-August-Universit\"{a}t G\"{o}ttingen.}}

\author{William D. Cochran, Michael Endl, Robert A. Wittenmyer
and Jacob L. Bean}
\affil{McDonald Observatory, The University of Texas at Austin, Austin TX 78712}
\email{wdc@astro.as.utexas.edu}

\begin{abstract}
We report the detection of two planetary mass companions to the solar-type
star HD\,155358.   The two planets have orbital periods of 195.0 and
530.3\,days, with eccentricities of 0.11 and 0.18.  The minimum masses for
these planets are 0.89 and 0.50 M$_J$ respectively.
The orbits are close enough to each other, and the planets are sufficiently
massive, that the planets are gravitationally interacting with
each other, with their eccentricities and arguments of periastron varying
with periods of 2300--2700 years.  
While large uncertainties remain in the orbital eccentricities,
our orbital integration calculations indicate that our
derived orbits would be dynamically stable for at least $10^8$ years.
With a metallicity [Fe/H] of -0.68, HD\,155358 is tied with
the K1III giant planet host star HD\,47536 for the lowest metallicity
of any planet host star yet found.
Thus, a star with only 21\% of the heavy-element content of our Sun was
still able to form a system of at least two Jovian-mass planets and have their
orbits evolve to semi-major axes of 0.6--1.2~AU.
\end{abstract}


\section{Introduction}
There is now strongly compelling evidence that planetary companions to main
sequence stars are found preferentially around metal-rich stars.
\citet{Go97,Go98a,Go98b,Go99} was the first to do a detailed comparison of the
chemical abundances of stars with and without planets, and to conclude that
there was a significant difference in these samples.  This finding has
subsequently been supported by a large number of investigations,
several of which included carefully selected control samples
\citep[e.g.][]{Re02,HeLu03,SaIsMa05,FiVa05,GrLi07}.
Much of the early work in this area concentrated on determining the
direction of causality in the observed correlation.
That is, does the presence of planets pollute the stellar convective region
causing an apparent increase in photospheric metallicity, or does the
physics of planet formation significantly favor a metal-rich environment?
This issue has been addressed by searching for patterns of elemental
enhancement in planet-host stellar photospheres that might result from
the metal-rich (or H- and He-poor) detritus of planet formation. 
So far, there is little compelling evidence that post-formation stellar
photospheric self-enrichment is responsible for
the observed correlation; a primordial explanation is most likely
\citep{Go06}.   Thus, searching for planets around stars with relatively
low metallicity is extremely important for our understanding of the physics
of planet formation.   We need to find the lowest metallicity stars around
which planets are able to form.  The radial-velocity survey of
\citet{SoToLa06} was the first to specifically target low metallicity
stars in order to understand in detail the low-$Z$ end of the dependence
of planetary system and formation on metallicity.

We present here the discovery of two planetary companions to the
star HD\,155358.  
This is one of the sample of low-metallicity stars that were
included in our HET planet survey \citep{CoEnMA04} in order to begin to
explore the low-metallicity tail of the heavy element distribution
of stars with planets.
Details of the observations are given in Section~\ref{observations}.
Orbital solutions for the planets are given in Section~\ref{orbits1}
and~\ref{orbits2}.  Section~\ref{genetics} describes our use of a
genetic algorithm to achieve a complete exploration of possible orbital
solutions.  The dynamical stability of our preferred orbital solution is
explored in Section~\ref{dynamics}.  
Detailed parameters of the host star are given in
Section~\ref{stellar_properties}, and the implications of this fascinating
system for planet formation theories are given in Section~\ref{discussion}.

\section{Observations}\label{observations}
All observations of HD\,155358 were made in queue scheduled mode
using the High Resolution Spectrometer \citep{Tu98} of the
9.2\,m Hobby-Eberly Telescope \citep{RaAdBa98}.
A {400\micron}  optical fiber which subtends 2\farcs0 on the sky
tracks the star across the focal plane of the telescope. 
Starlight then passes through a molecular iodine absorption
cell stabilized at a temperature of 70C.
The dense narrow I$_2$ spectrum which is superimposed on the stellar
spectrum enables us to model the instrumental point spread function
\citep{VaBuMa95} and provides the velocity metric for precise radial
velocity measurement \citep{BuMaWi96,EnKuEl00}.
A {250\micron} spectrograph entrance slit then gives a spectral resolving
power of $R = \lambda/\delta\lambda = 60,000$.
The spectrum is recorded on a mosaic of two $2048 \times 4100$ pixel E2V
CCDs, which samples the spectrum at $\approx 4$ pixels per resolution
element.
The cross-disperser is set to obtain the spectrum between 4076{\AA} and
7838{\AA}.  A small gap between the two CCDs falls at 5936{\AA}, allowing most
of the I$_2$ absorption spectrum to be recorded on the ``blue'' CCD chip.

We obtained a total of 71 high signal/noise radial velocity observations
of HD\,155358 between June 2001 and March 2007.  
All spectra were processed using standard IRAF routines,
and velocities were computed using our high-precision radial velocity code.
Table~\ref{tab:velocities} gives the relative radial velocities for
HD\,155358.  The observation times and velocities have
been corrected to the solar system barycenter.   The uncertainty
$\sigma$ for each velocity in the table is an {\itshape internal} error
computed from the variance about the mean of the velocities from each of
the 480 small chunks into which the spectrum is divided for the velocity
computation.   Thus, it represents the relative
uncertainty of one velocity measurement with respect to the others, based
on the quality and observing conditions of the spectrum.  This uncertainty
does not include other intrinsic stellar sources of uncertainty, nor any
unidentified sources of systematic errors.
\begin{deluxetable}{rr@{.}lr}
\tablecolumns{4}
\tablecaption{Relative Velocities for HD\,155358 \label{tab:velocities}}
\tablehead{
\colhead{BJD}&
\multicolumn{2}{c}{Velocity}&
\colhead{$\sigma$} \\
\colhead{-2\,400\,000} &
\multicolumn{2}{c}{m\,s$^{-1}$} &
\colhead{m\,s$^{-1}$}}
\startdata
52071.904837   &   19&49  & 3.28 \\
52075.886827   &   24&07  & 3.17 \\
52076.889856   &   25&29  & 2.93 \\
52091.846435   &   31&10  & 3.08 \\
52422.937143   &    3&70  & 3.43 \\
53189.839130   &  -30&67  & 3.28 \\
53205.795629   &  -20&54  & 3.08 \\
53219.753215   &  -16&15  & 3.35 \\
53498.776977   &   24&06  & 3.72 \\
53507.957634   &   41&99  & 4.30 \\
53507.962870   &   41&12  & 3.81 \\
53511.953195   &   41&06  & 2.88 \\
53512.939314   &   34&47  & 3.29 \\
53590.733229   &  -13&27  & 2.86 \\
53601.700903   &   -3&25  & 2.77 \\
53604.701932   &   -3&05  & 2.97 \\
53606.702997   &   -5&41  & 3.20 \\
53612.684038   &    3&44  & 2.61 \\
53625.645464   &   29&05  & 2.65 \\
53628.624309   &   28&41  & 3.61 \\
53629.619464   &   23&41  & 2.85 \\
53633.616275   &   20&48  & 2.83 \\
53755.042134   &  -35&55  & 4.38 \\
53758.042744   &  -41&84  & 3.70 \\
53765.035558   &  -30&94  & 4.02 \\
53774.021743   &  -45&17  & 4.33 \\
53779.979871   &  -26&31  & 4.27 \\
53805.930694   &   -5&23  & 3.16 \\
53808.892822   &   -4&14  & 4.28 \\
53769.026048   &  -42&02  & 3.56 \\
53866.964277   &   30&26  & 3.87 \\
53869.954988   &   29&05  & 3.55 \\
53881.945116   &   43&83  & 2.66 \\
53889.682359   &   47&51  & 2.83 \\
53894.672403   &   42&03  & 2.80 \\
53897.895406   &   34&74  & 2.39 \\
53898.685344   &   27&26  & 2.75 \\
53899.892253   &   36&21  & 2.91 \\
53902.657932   &   17&67  & 2.53 \\
53903.906452   &   35&85  & 2.87 \\
53904.662338   &   26&89  & 3.03 \\
53905.860234   &   27&26  & 3.10 \\
53907.629415   &   28&25  & 2.75 \\
53908.876910   &   23&17  & 3.05 \\
53910.639802   &   26&13  & 3.61 \\
53911.845252   &   17&72  & 2.70 \\
53912.636138   &   15&77  & 3.03 \\
53917.641443   &    9&13  & 3.12 \\
53924.818858   &   14&17  & 3.21 \\
53925.807994   &   10&44  & 2.60 \\
53926.824657   &   -0&29  & 2.83 \\
53927.822039   &    9&12  & 2.98 \\
53936.790961   &   -7&51  & 2.98 \\
53937.790352   &  -12&14  & 6.01 \\
53937.804652   &   -0&17  & 2.83 \\
53941.762397   &  -14&31  & 2.78 \\
53943.766914   &  -11&76  & 2.83 \\
53954.756895   &  -10&92  & 2.49 \\
53956.734767   &  -15&38  & 2.47 \\
53958.747552   &  -11&09  & 2.50 \\
53960.714843   &   -9&99  & 2.52 \\
53966.708140   &  -11&85  & 2.34 \\
53971.697987   &  -12&96  & 2.51 \\
53985.651466   &   12&67  & 3.81 \\
53988.637456   &    6&17  & 2.83 \\
53990.639454   &   19&80  & 3.29 \\
53993.644925   &   20&57  & 3.27 \\
54136.010987   &  -17&45  & 4.47 \\
54137.005715   &   -9&57  & 3.98 \\
54165.933759   &  -30&96  & 3.73 \\
54167.924067   &  -24&81  & 3.85\enddata
\end{deluxetable}

\section{Orbital Solution}
\subsection{The First Planet}\label{orbits1}
A periodogram of the HET velocities of HD\,155358 show a very strong
peak at $\sim$195\,days, with substantial additional power in the
300-500 day range.
The \citet{Sc82} false-alarm probability of this 195-day signal is 
$4.5\times10^{-10}$.   We first fit a Keplerian orbit to the data using the
{\itshape GaussFit} generalized least squares software of \citet{JeFiMA88}.
This planet, with a period of 193.8\,d, eccentricity of 0.16, and
K~velocity of 31.0\,m\,s$^{-1}$ has a minimum mass of
$M \sin i = 0.79\,$M$_{\rm J}$ and a semimajor axis of 0.63\,AU.
The RMS dispersion of the data around the orbital solution is
10.2\,m\,s$^{-1}$, and the reduced chi-squared
$\chi^2_{\nu} = 9.67$.   This goodness-of-fit parameter is significantly
higher than would be expected based, as it is based purely on the internal
errors quoted in Table~\ref{tab:velocities}.
\citet{Wr05} gives a median ``jitter'' of 4.4\,m\,s$^{-1}$ for
main-sequence stars similar to HD\,155358.  Wright also
mentions that subgiants typically have jitters of about 5\,m\,s$^{-1}$,
as do blue stars (with B-V $<$ 0.6). 
HD 155358 appears to fall into both categories (cf.
section~\ref{stellar_properties}, with our derived age of 10 Gyr
and Hipparcos B-V=0.545.
While some of the additional scatter might be explained as intrinsic stellar
velocity variability expected in this star, there is probably some
additional  source for much of the scatter.  
Thus, we have searched carefully for additional planets in the system.

\subsection{The Second Planet}\label{orbits2}
\begin{figure*}[ht]
\plotone{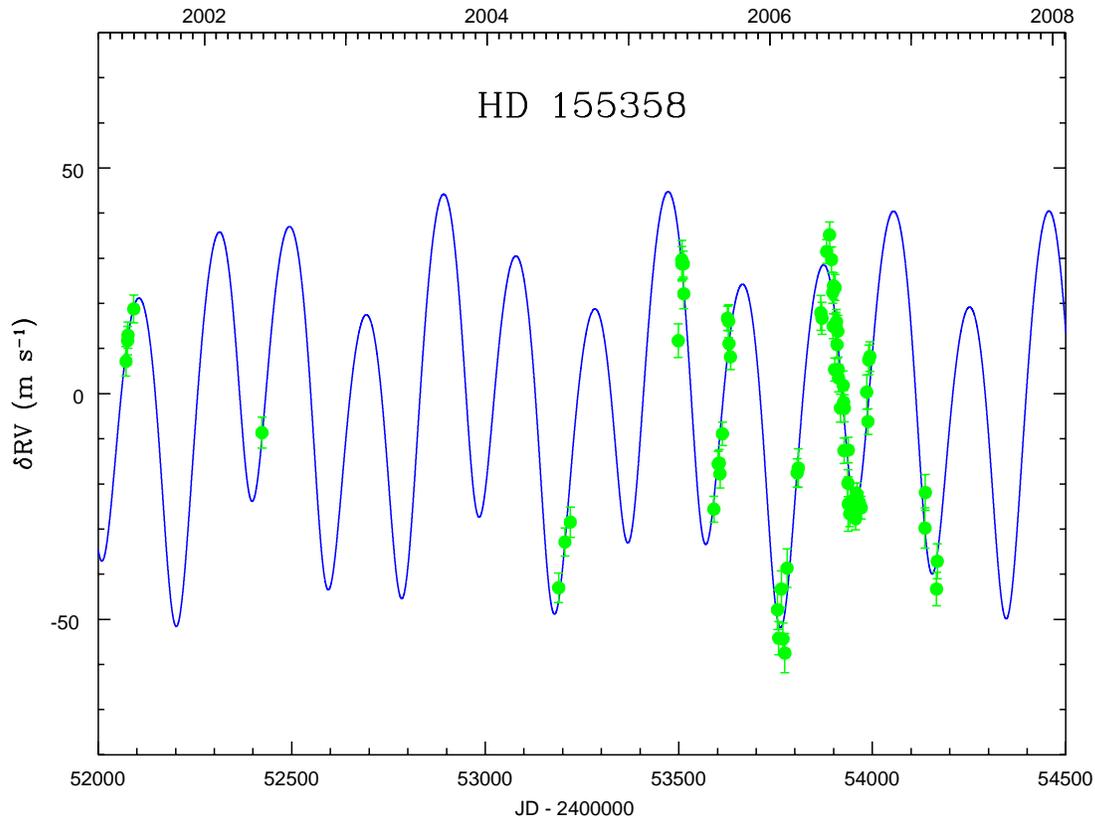}
\caption{The HET data velocities of HD\,155358 with the
best double-planet fit (solid line).
The RMS scatter of the data around the fit is 6.0\,m\,s$^{-1}$.
\label{fig:orbits}}
\end{figure*}

\begin{figure*}[ht]
\plotone{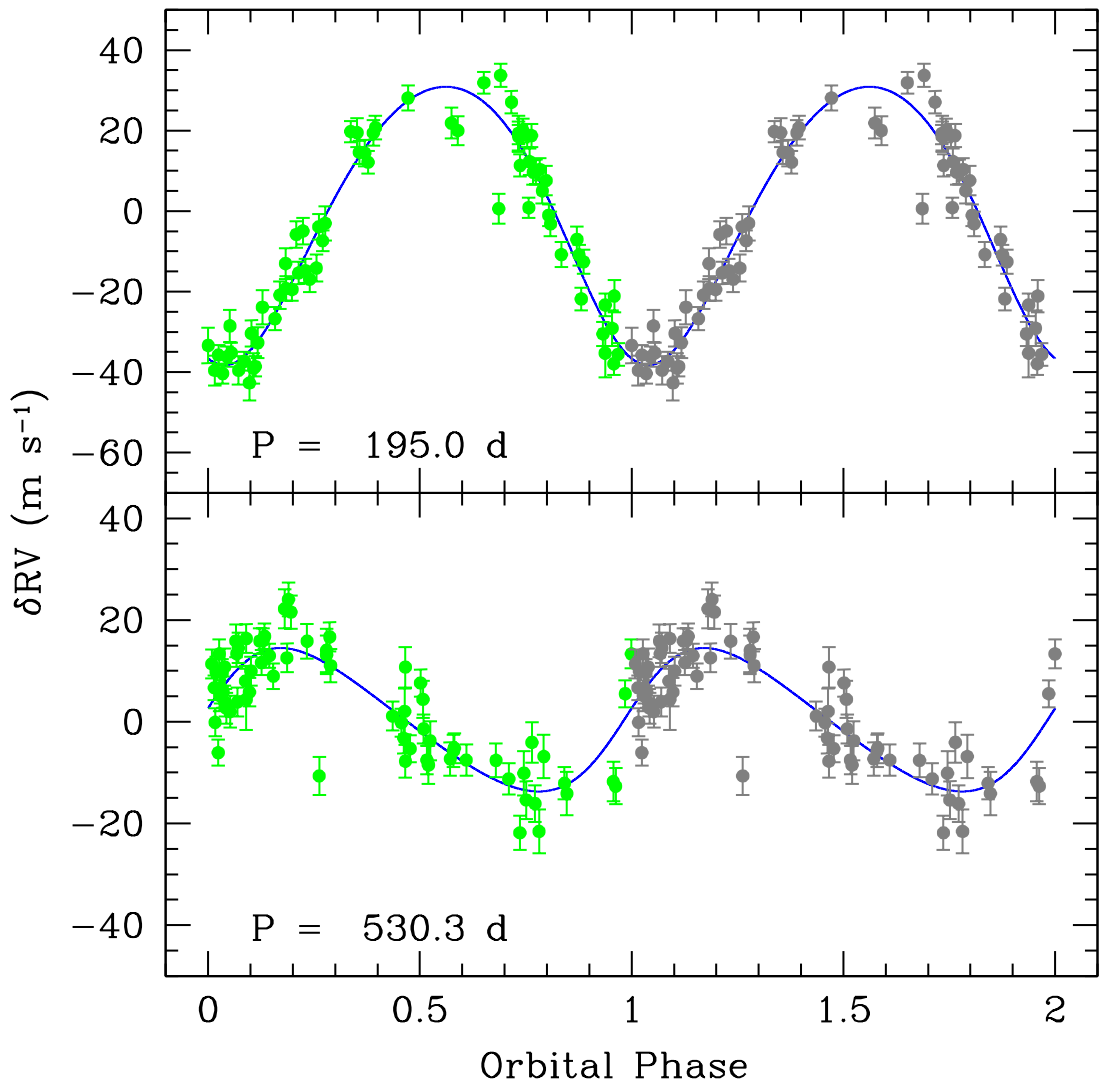}
\caption{The best fit double-planet Keplerian orbit solution to the
Hobby-Eberly Telescope data for HD\,155358.  The upper panel shows
the phased inner planet (HD\,155358\,b) orbit, along with the observed data
which have had the outer planet (HD\,155358\,c) orbit removed. 
The lower panel shows the phased HD\,155358\,c orbital solution,
along with the observed data which have had the the HD\,155358\,b orbit
removed.  Two cycles are shown, with the second cycle in grey for clarity.
The rms scatter of the data around this double-planet fit
is 6.0\,m\,s$^{-1}$. \label{fig:planet2phased}}
\end{figure*}

A Lomb-Scargle periodogram of the residuals around the single-planet orbital
fit shows very significant power over a broad range of periods
from 200 to 800 days, with the strongest peak around 530\,days.  
The false-alarm probability of this power at 530\,days is less
than $10^{-6}$.   Smaller peaks around 330 days show FAP slightly less
than $10^{-3}$.  Thus, we investigated the possibility
of a second long-period planet HD\,155358\,c.
We used {\itshape GaussFit} to compute simultaneous least-squares solutions of
double-Keplerian orbits to the observed velocities.  
We investigated periods around both 330 days and 530 days.
While we were able to find formal Keplerian solutions for both periods,
the solutions with P$_{\rm c} \sim 530$\,days gave significantly better fits
than did shorter period second planets.
When we couple this with the results of our genetic algorithm orbital fits
(section~\ref{genetics}) and dynamical calculations (section
\ref{dynamics}), we have concluded that the 530 day period is
correct for HD\,155358\,c.  This solution is given in
Table~\ref{tab:planet2} and is shown in Figure~\ref{fig:orbits}.  
The uncertainties reported in Table~\ref{tab:planet2} were generated by
{\itshape GaussFit}
from a maximum likelihood estimation that is an approximation to a Bayesian
maximum a posteriori estimator with a flat prior \citep{Je90}.
The observed HET velocity residuals to the
HD\,155358\,c fit, phased to the period of HD\,155358\,b are shown in
the upper panel of Figure~\ref{fig:planet2phased}, and vice-versa in the
lower panel.
\begin{deluxetable}{lr@{$\pm$}l}
\tablecolumns{3}
\tablecaption{HD\,155358 Double Planet Keplerian Solution
\label{tab:planet2}}
\tablehead{
\colhead{Parameter} &
\multicolumn{2}{c}{Value}}
\startdata
P$_{\rm b}$ & 195.0 & 1.1 days \\
T$_{0 \rm b}$ & 2453950.0 & 10.4 BJD \\
K$_{\rm b}$     & 34.6 & 3.0 m\,s$^{-1}$ \\
e$_{\rm b}$    & 0.112 & 0.037 \\
$\omega_{\rm b}$ & 162 & 20 degrees \\
\hline
P$_{\rm c}$ & 530.3 & 27.2 days \\
T$_{0 \rm c}$ & 2454420.3 & 79.3 BJD \\
K$_{\rm c}$     & 14.1 & 1.6 m\,s$^{-1}$ \\
e$_{\rm c}$     & 0.176 & 0.174 \\
$\omega_{\rm c}$ & 279 & 38 degrees \\
\hline
$M_{\rm b} \sin i $ & 0.89 & 0.12 $M_J$ \\
a$_{\rm b}$ & 0.628 &  0.020 AU\\
$M_{\rm c} \sin i $ & 0.504 & 0.075 $M_J$ \\
a$_{\rm c}$ & 1.224 &  0.081 AU\\
RMS &  \multicolumn{2}{c}{6.0 m\,s$^{-1}$} \\
\enddata
\end{deluxetable}

In fitting a Keplerian orbit to observed data,
the eccentricity is often the least well-determined of the orbital
elements, and is the one element that is most often able to absorb any
additional signals or periodicities that may be in the data.   For example,
in comparing the HD\,155358\,b single-planet orbit fit with the elements of 
HD\,155358\,b in the two-planet fits, the most significant changes in the
orbital elements of HD\,155358\,b are that the eccentricity dropped somewhat
and the K value increased.   The uncertainty in the eccentricity of
HD\,155358\,c is rather large.  This is probably due to some regions of
sparse phase coverage, particularly in the phase intervals 0.30-0.45 and
0.8-1.0.  The rms scatter of the observations around this two-planet fit is 
6.0\,m\,s$^{-1}$.   This rms is quite consistent with a stellar internal
``jitter'' of 4.5 to 5.0\,m\,s$^{-1}$ and our typical internal observed
velocity uncertainty of 2-4\,m\,s$^{-1}$.
If we add a stellar ``jitter'' of 5\,m\,s$^{-1}$ in quadrature with the
internal errors given in Table~\ref{tab:velocities},
we then get a reduced chi-squared for our two-planet fit of 1.15.

\subsection{Genetic Algorithm Investigation \label{genetics}}
In order to understand fully the nature of the orbit of the second planet in
this system, we used a standard genetic algorithm to explore further the
$\chi^{2}$-landscape of a 2 planet solution and to find possible
additional $\chi^2$ minima. The initial orbital parameters were distributed
randomly over a certain range of start values and during each iteration
these parameters were randomly mutated. The solutions ``evolve'' by using
$\chi^{2}$ as the environmental selection mechanism. Solutions that
improve $\chi^{2}$ or at least do not degrade it by a certain threshold
are allowed to survive and to multiply. Solutions that are above this
threshold (e.g. $\chi^{2} < \chi^{2}_{\rm best}$ + 0.1\%), and are not
able to move back into this $\chi^{2}$ range, are terminated after a few
generations (typically 6 generations). Repeating this process many times
(each time with new randomly selected starting values) allowed us to
probe a much larger parameter space than is usually possible with
standard orbital fit programs.

For HD\,155358 we performed $>10,000$ trial runs, allowing the period of
HD\,155358\,c to vary between 250 and 2100 days (the time span of
observations). We constrained the period of HD\,155358\,b to the range of
185 to 200 days and its eccentricity to $e_{\rm b} < 0.5$.

Three $\chi^{2}$ minima were found by the genetic algorithm: the
first with periods for HD\,155358\,c around 326 days, the second around
with P$_{\rm c}$ 530 days and a third one with P$_{\rm c}$ around 1540 days.
The parameters of the solutions found for the first two
minima coincide with the solutions found by {\itshape GaussFit}.
However, the third minimum at 1540 days was intriguing, as it
represents another possible period for HD\,155358\,c.  
A detailed examination of the solutions in this minimum revealed that
their eccentricities were all in the range of $0.835$ to $0.86$.  
With such a high eccentricity the orbits of the
two planets would cross and the system would quickly disrupt itself.
It appears that this 1540\,day solution represents
a good formal fit for a physically impossible orbital configuration.
In order to further rule out the 1540 days as the true period of
the outer planet we performed 1000 additional trials in the period
range of 1500 to 1600 days for HD\,155358\,c and using an upper
limit on $e_{\rm c}$ of 0.6.
This time no solutions was found with comparable low $\chi^{2}$ values.
We thus conclude that the 530 day period is the most likely period for the
second planet.

\subsection{Dynamical Stability Calculations \label{dynamics}}

While the orbits given in Table~\ref{tab:planet2} represent the least-squares 
Keplerian orbit solution to the observations, they may not necessarily
represent physically realistic orbits for two planets in this system.  
The inclusion of the second planet HD\,155358\,c drops the eccentricity of
the first planet HD\,155358\,b slightly, but the new outer planet
solution has a somewhat non-zero eccentricity itself. 
While the orbits for this solution do not cross each other, they do have
sufficiently close approaches that the planets may well be interacting
dynamically.

To investigate the orbital stability of the two-planet Keplerian orbital
solution for the HD\,155358 system, we conducted dynamical simulations using
the N-body integrator SWIFT\footnote{SWIFT is publicly available at
http://www.boulder.swri.edu/$\sim$hal/swift.html.}.  A full description of
SWIFT's capabilities is given by \citet{LeDu94}.  We adopted the
stellar mass $M_*$=0.87\,M$_{\sun}$ as discussed below in
Section~\ref{stellar_properties}, and the parameters of the two planets were
those given in Table~\ref{tab:planet2}.
The planets were assumed to be coplanar, and the planet masses were taken to be
the minimum values (i.e. $\sin i=1$).  All simulations which used
shorter period in the range of 300 days discussed above in
Section~\ref{genetics}  for the outer HD\,155358\,c planet resulted
in ejection of one or both planets within $10^5$~yr.
Thus, we reject all solutions with P$_{\rm c}$ in the 300\,day range.
The longer period (P$_{\rm c}=530.3$ days) for HD\,155358\,c is favored
dynamically, and was used in all subsequent simulations.

\begin{figure*}[ht]
\plotone{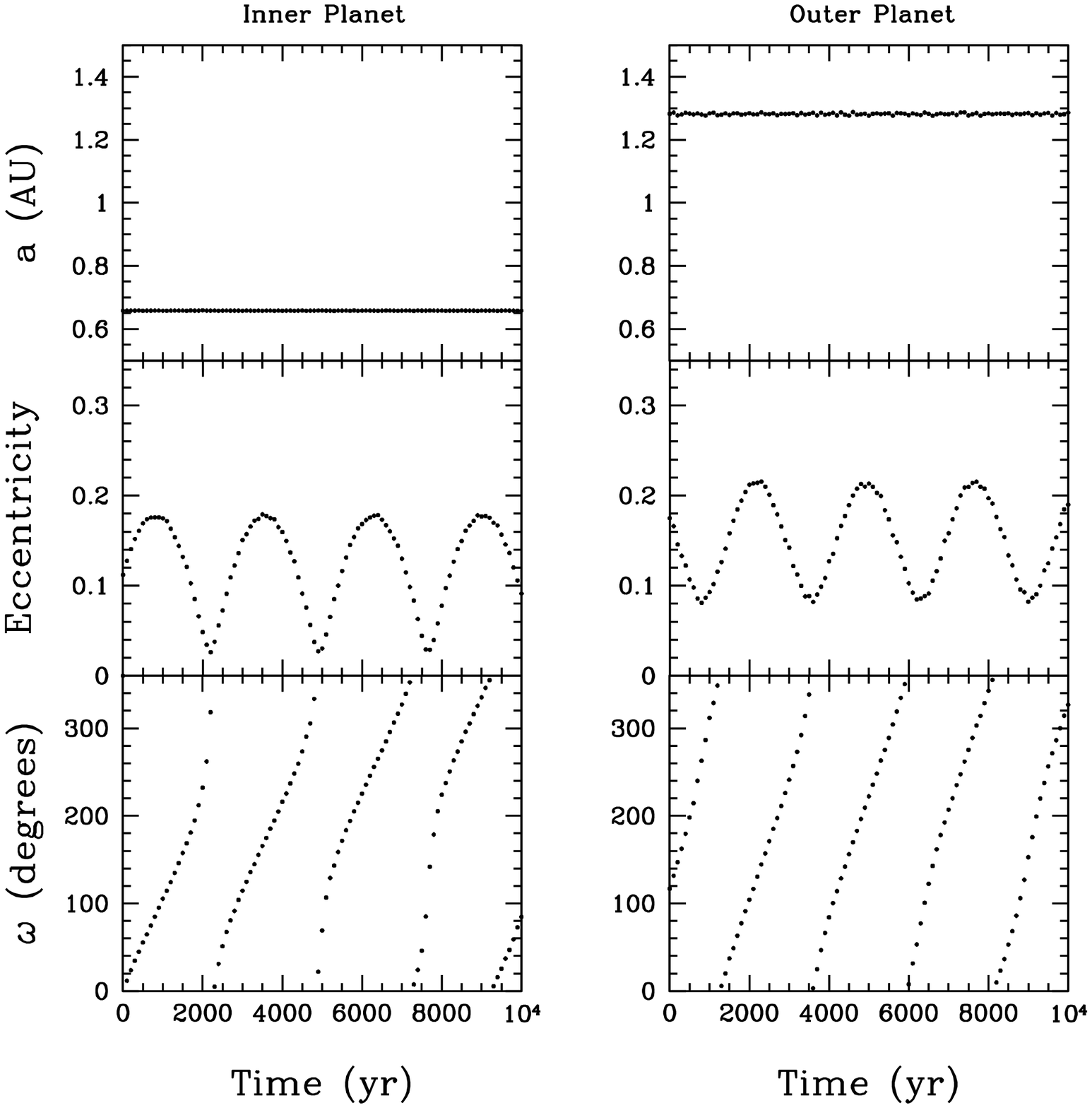}
\caption{ Results of dynamical simulations of the HD\,155358 system.  The
evolution of semimajor axis (a), eccentricity, and the argument of periastron
($\omega$) are shown for a typical $10^4$ year interval. 
The simulation was run for a total of $10^8$ years. 
The two planets exchange eccentricities on a timescale of about 2700~yr,
and $\omega$ circulates with a period of about 2300~yr.
The semimajor axes of both planets are stable throughout the simulation. 
These computations were for a star of mass 0.87\,M$_{\sun}$, and planet
masses fixed at the minimum mass ($\sin i = 1$).   However, similar
results are obtained for $\sin i\gtrsim 0.4$.  \label{fig:dynamics} }
\end{figure*}

The three-body system (star and two planets) was integrated for a total of
$10^8$ yr.   Initial orbits for the planets were taken from
Table~\ref{tab:planet2}.
The system remained stable for the duration of the simulation.  As shown
in Figure~\ref{fig:dynamics}, the two planets exchanged eccentricities on a
relatively short timescale, with a period of approximately 2700~yr. 
The eccentricity of planet b varied between 0.02 and 0.18, while the
eccentricity of planet c varied between 0.08 and 0.22.  The
argument of periastron $\omega$ for both planets circulated with a period
of about 2300~yr.   The two planets spend most of their time with
$\left|\omega_{\rm b} - \omega_{\rm c}\right|$ between $120\arcdeg$ and
$240\arcdeg$.

Noting the significant interaction between the two planets in the
minimum-mass case, we tested the effect of inclination of the entire system
by adjusting the planetary masses and repeating the simulations. 
More massive planets (sin~$i<1$) should interact more strongly with each
other, and thus may be less stable.  Systems
with $\sin i\gtrsim 0.33$ ($i\gtrsim$20\arcdeg) remained stable for the
duration of the tests ($10^7$~yr).  These results indicate that the
dynamical stability evidenced in Figure~\ref{fig:dynamics} does not
require the special circumstance of a nearly edge-on inclination and
thus minimum-mass planets.
Even though the two planets are definitely interacting dynamically, the
timescale of this interaction is long compared with the span of our
observations.   Thus, our two-planet least-squares Keplerian solution given
in Section~\ref{orbits2} is a valid approximation.

Throughout the dynamical simulations, we have assumed that the two planets
are in coplanar orbits.  It is, however, conceivable that the planetary
orbits are inclined with respect to each other.  We have conducted further
tests with mutually inclined planetary orbits to determine the maximum
mutual inclination for which the system remained stable. 
For the purposes of these experiments, the
initial inclination of the inner planet was set to $0\arcdeg$ and that of
the outer planet was assigned a grid of starting values ranging from
$5\arcdeg$ up to $50\arcdeg$.  The planetary masses were again assumed to be
the minimum values, and the systems were integrated for $10^7$~yr.
Systems with mutual inclinations less than $45\arcdeg$ remained stable for
the duration of the simulations.  For higher values of the inclination
between the two planets, the eccentricities of both planets experienced
chaotic variations resulting in the ejection of the outer planet within
$10^6$~yr ($45\arcdeg$) and $2\times 10^5$~yr ($50\arcdeg$).

\section{Stellar Properties} \label{stellar_properties}
We measured the photospheric iron abundance, [Fe/H], of HD\,155358 by analyzing
the ``template'' spectrum used in the radial velocity analysis. We assumed
that [Fe/H] was an effective proxy for the general photospheric
metallicity, [M/H]. Our analysis method is the same as described by
\citet{BeSnHa06} for solar-type stars. We briefly describe the technique
here, and refer the reader to that paper for a complete description.

We fit synthetic spectra to the profiles of 30 Fe I lines in the observed
spectrum. We generated the synthetic spectra with an updated version of
the plane-parallel, local thermodynamic equilibrium (LTE), stellar
analysis computer code MOOG \citep{Sn73}. We assumed astrophysical log
\textsl{gf} values for the Fe I lines, which were determined by fitting a
solar spectrum and assuming the solar iron abundance log
$\epsilon$(Fe)$_{\sun}$ = 7.45. Our analysis was therefore differential to
the sun.

We also adopted model atmospheres computed with the general-purpose
stellar atmosphere code PHOENIX \citep[version 13,][]{HaAlBa99} for our
analysis. The model atmosphere effective temperature, $T_{eff}$, was
constrained using the \{$(\bv)$, [Fe/H]\} -- $T_{eff}$ relationship of
\citet{RaMe05}. The surface gravity, log \textsl{g}, was constrained by
interpolating the \citet{BeBrCh94} evolutionary isochrones for a given
$M_{V}$, $T_{eff}$, and [Fe/H]. We took the needed $V$ magnitude, $(\bv)$
color, and parallax from the \textit{Hipparcos} catalog \citep{PeLiKo97}.

We determined [Fe/H], microturbulence $\xi$, and macroturbulence $\eta$ for
HD\,155358 by fitting the line profiles in the observed
spectrum simultaneously. As the constraints on the stellar $T_{eff}$ and
log \textsl{g} described above are both dependent on [Fe/H], these
parameters were also varied simultaneously. We fit the observed spectrum
by using an adaptation of the ``Marquardt'' $\chi^{2}$ minimization
algorithm \citep{MA63,PrFlTe86}. The uncertainty in the determined [Fe/H]
was calculated from the scatter in the abundances determined for each line
individually and deviations due to the uncertainties in the adopted
$T_{eff}$ and log \textsl{g}.

In addition to the standard spectroscopic parameters, we have also
estimated the mass and age of HD\,155358 by interpolating the \citet{BeBrCh94}
evolutionary isochrones as we did to constrain the log \textsl{g}. The
parameters that we determined for HD\,155358 are given in
Table~\ref{tab:ourparameters}. Table ~\ref{tab:compareparameters} lists
the literature values of the stellar parameters for comparison. Most
notably, we found [Fe/H] = -0.68 $\pm$ 0.07. This is in very good
agreement with the literature values, which have a range -0.72 $\leq$
[Fe/H] $\leq$ -0.60, and median value [Fe/H] = -0.67.  The derived value
of $\log g$, coupled with the mass, age, and effective temperature of
HD\,155358 indicate that the star has evolved off the main sequence.
\begin{deluxetable}{lr@{$\pm$}l}
\tablecolumns{2}
\tablecaption{Derived Stellar Parameters for HD\,155358
\label{tab:ourparameters}}
\tablehead{
 \colhead{Parameter} &
 \colhead{Value}
}
\startdata
 $T_{eff}$       &  5760 & 101 K            \\
 log \textsl{g}  &  4.09 & 0.10 (cgs)       \\
 ${\rm [Fe/H]}$          & -0.68 & 0.07             \\
 $\xi$           &  1.55 & 0.15 km s$^{-1}$ \\
 $\eta$          &  1.89 & 0.20 km s$^{-1}$ \\
 $M_{\star}$     &  0.87 & 0.07 $M_{\sun}$  \\
 Age             &  \multicolumn{2}{c}{$\sim$ 10 Gyr} \\
\enddata
\end{deluxetable}

\begin{deluxetable}{cccccc}
\tablecolumns{6}
\tablecaption{Stellar Parameters for HD\,155358 from Other Sources
\label{tab:compareparameters}}
\tablehead{
 \colhead{$T_{eff}$} &
 \colhead{log \textsl{g}} &
 \colhead{[Fe/H]} &
 \colhead{Mass} &
 \colhead{Age} &
 \colhead{Source} \\
 \colhead{(K)} &
 \colhead{(cgs)} &
 \colhead{} &
 \colhead{($M_{\sun}$)} &
 \colhead{(Gyr)} &
 \colhead{}
}
\startdata
 5760   &  4.09   &  -0.68  &  0.87   &  10.1   & 1  \\
 5870   &  4.19   &  -0.67  &  0.75   & \nodata & 2  \\
 5868   &  4.19   &  -0.67  & \nodata & \nodata & 3  \\
 5914   &  4.09   &  -0.61  & \nodata & \nodata & 4  \\
\nodata & \nodata & \nodata &  0.84   &  10.2   & 5  \\
 5860   &  4.10   &  -0.60  & \nodata & \nodata & 6  \\
 5818   &  4.09   &  -0.67  &  0.89   &  12.7   & 7  \\
 5888   & \nodata &  -0.69  &  0.83   &  13.8   & 8  \\
\nodata & \nodata & \nodata & \nodata &  15.7   & 9  \\
 5808   & \nodata &  -0.72  &  0.89   &  11.9   & 10 \\
\enddata
\tablerefs{(1) This Paper, (2) \citet{LaHeEd91}, (3)
\citet{EdAnGu93}, (4) \citet{GrCaCa96}, (5) \citet{NgBe98},
(6) \citet{ThId99}, (7) \citet{ChNiBe01}, (8) \citet{FeHoHu01},
(9) \citet{LaRe04}, (10) \citet{NoMaAn04}}
\end{deluxetable}

\section{Discussion}\label{discussion}
The planetary system around HD\,155358 is quite remarkable in that it
contains two Jovian-mass planets in relatively short-period orbits
around one of the most metal-poor planet-host star yet found.
Thus, this system is crucial to our understanding of the
physics of planetary system formation and early dynamical evolution.

The final configuration of any planetary system results from a competition
of several different physical processes, each occurring on its own independent
timescale.   The planets themselves must form from the remnant disk of
material left over from star formation.   This must happen before the
circumstellar disk is dissipated.   After planet formation, the system must
evolve into the final state in which we detect it.   The dependence of each
of these processes on the metallicity of the material from which the
entire system forms then determines the range of allowed planetary systems
as a function of metallicity.

\citet{MaShOo07} investigated the circumstances under which stars of
various metallicity and mass could form planets within the framework of
both the core-accretion model and the disk-instability model.
They derive a low metallicity threshold for planet formation within the
core accretion model of [Fe/H] = -0.85 for a disk of 5 times the mass of
the minimum mass solar nebula (MMSN) and [Fe/H] = -1.17 for a disk of 10
times the MMSN.  They also showed that the disk instability model has
essentially no dependence on stellar metallicity, but would require a disk
mass of at least 10 times the MMSN to form these planets.
Thus, the two giant planets we have discovered in the HD\,155358 system
could have formed from either formation mechanism. 
In either model they require a disk which was
substantially more massive than the minimum-mass solar nebula.

\citet{IdLi04II,IdLi05II} investigated the dependence
of the formation of planetary systems on stellar metallicity
within the core-accretion model.   The motivation for their model was to
explain the observed correlation between high stellar metallicity and
planetary systems easily detected by RV techniques.   The basic idea is
that in systems of high metallicity, the dust surface density in the
protoplanetary disk will be enhanced.   This will lead to planetary cores
being formed on a much shorter timescale.  \citet{IdLi05II} derived a
planetary core mass at time $t$ (prior to depletion of the feeding zone)
that is proportional to $10^{3Z}$, where $Z$
is the stellar metallicity ($10^{\rm [Fe/H]_{\star}} Z_{\sun}$).
For a giant planet to grow, a core must achieve a mass above a critical
value of about 10-20$M_{\oplus}$ before the gas disk is dissipated.
If the core does reach a super-critical mass while there is still significant
gas content in the disk, it will rapidly accrete gas and grow to become a
gas-giant planet.   If the core remains sub-critical through gas disk
dissipation, then the planetary core will remain a Neptune or a
super-Earth.
Since the disk dissipation has a much weaker (if any) dependence on stellar
metallicity, gas-giant planet formation around low $Z$ stars is
significantly more difficult than around stars of solar or higher
metallicity.
Unless the circumstellar disk is rather massive to begin with,
there simply isn't time to grow critical-mass cores before the gas disk
dissipates.   Thus, one might expect a critical lower metallicity threshold
for the formation of gas giant planets.  The value of this threshold would
depend on the physics governing the overall mass of the disk, which are
poorly understood.

The third relevant timescale for determining the final configuration of a
planetary system is that of the planetary system dynamical
evolution, which transforms the system from the one which formed the
planets into the system which we detect today.
There are two basic physical processes governing post-formation dynamical
evolution.  These are tidal migration prior to disk dissipation, and
planet-planet or planet-planetessimal scattering following disk dissipation.  
Only tidal migration
can have any significant dependence on stellar metallicity.  The disk
evolution is driven by the disk viscosity.  In type II migration,
in which the planet has opened a gap in the disk as a result of
depleting the local gas in the disk from runaway gas accretion,
the planet is locked into the viscous evolution of the disk.
\citet{IdLi05II} give a disk migration timescale that depends linearly on
the disk viscosity,~$\alpha$.    The dependence of $\alpha$ on metallicity
is not known, and most models assume a constant $\alpha$ of order $10^{-4}$
for all disks.
Thus, it appears that once a system of several planets is able to form in a
disk, the subsequent dynamical evolution of that system is probably
reasonably independent of~$Z$.   Detailed modeling of the dependence of
disk evolution and type~II migration on $Z$ is needed.
If the dynamical evolution of the planetary system is due to tidal migration
in the disk, this migration must still occur before the disk is dissipated.
If the final configuration of the HD\,155358 system is due to
post-formation planet-planet scattering, or dynamical interactions of the
planets with remnant planetessimals, then we would not necessarily expect
any significant dependence of the process on the stellar metallicity.
For low metallicity systems, the theoretical challenge is to form the
planets of the required mass before the disk dissipates.

\acknowledgments
This material is based on work supported by the National Aeronautics and
Space Administration under Grants NNG04G141G and NNG05G107G issued through
the Terrestrial Planet Finder Foundation Science program.
The dynamical simulations used in this work were carried out at the
Texas Advanced Computing Center (TACC).
The Hobby-Eberly Telescope (HET) is a joint project of the University of
Texas at Austin, the Pennsylvania State University, Stanford University,
Ludwig-Maximilians-Universit\"{a}t M\"{u}nchen,
and Georg-August-Universit\"{a}t G\"{o}ttingen.
The HET is named in honor of its principal benefactors,
William P. Hobby and Robert E. Eberly.

Facilities: \facility{HET}

\newcommand{\noopsort}[1]{}
  \newcommand{\printfirst}[2]{#1} \newcommand{\singleletter}[1]{#1}
  \newcommand{\switchargs}[2]{#2#1}

\end{document}